\documentclass{ws-procs975x65}

\usepackage{graphicx,epsfig, amsmath,amssymb,amsfonts}

\begin{document}

\title{Lorentz symmetry from Lorentz violation in the bulk}

\author{Orfeu Bertolami$^{1,2}$ and Carla Carvalho$^{1,2}$}

\address{$^1$ Departamento de F\'{\i}sica, Instituto Superior
  T\'ecnico, Avenida Rovisco Pais 1, 1049-001 Lisboa, Portugal}
\address{$^2$ Centro de F\'{\i}sica dos Plasmas, Instituto Superior
  T\'ecnico, Avenida Rovisco Pais 1, 1049-001 Lisboa, Portugal}

E-mail addresses: orfeu@cosmos.ist.utl.pt, ccarvalho@ist.edu

\begin{abstract}
We consider the mechanism of spontaneous symmetry breaking of a bulk
vector field to study signatures of bulk dimensions invisible to the
standard model confined to the brane.
By assigning a non-vanishing vacuum expectation value to the vector
field, a direction is singled out in the
bulk vacuum, thus breaking the bulk Lorentz symmetry. We present the
condition for induced Lorentz symmetry on the brane, as
phenomenologically required, noting that it is related to the value
of the observed cosmological constant.
\end{abstract}

\bodymatter

\section{Introduction}\label{intro}

Braneworld scenarios have changed our view of the extra
dimensions. The various models predict that gravity in our braneworld
can exhibit significant deviations from that described by Einstein's
general relativity.
In particular, in string theory inspired scenarios which assume that
the background bulk spacetime is anti-de Sitter,
it is possible to cancel out any 4-dimensional brane contribution to
the cosmological constant (see e.g. [1] and references
therein). Although not on its own a solution for the cosmological
constant problem, it is suggestive that braneworld scenarios might be an
important feature of a consistent description of the world.

It is therefore relevant to investigate the implications of the
braneworld scenarios to the formulation of
fundamental symmetries, another fundamental ingredient of the
physical description.
Lorentz symmetry, being from the phenomenological point of view one
of the most well and stringently tested symmetries of physics,
is particularly suitable to test the relation between bulk and
brane symmetries as a possible signature for the existence of extra
dimensions.

The possibility of violation of Lorentz invariance
has been extensively discussed in the recent literature (see
e.g. [4]) and in particular its astrophysical implications have been
studied\cite{Sato}. Furthermore, a connection between the cosmological
constant and the violation of Lorentz invariance has been conjectured in
the context of the string field theory\cite{Bertolami3}.

In this contribution we report on a recent study whose motivation was to
understand the way spontaneous Lorentz violation in the bulk is related to
Lorentz symmetry on the brane\cite{BeCarvalho06}.
We consider a bulk vector field coupled non-minimally to the graviton
which, upon acquiring a non-vanishing expectation
value in the vacuum, introduces spacetime anisotropies in the
gravitational field equations through the coupling with the
graviton\cite{Kostelecky3}.
After deriving the
equations of motion in the bulk, we project them parallel and
orthogonal to the surface of the brane.
The brane is assumed to be a distribution of
$Z_{2}$--symmetric stress-energy about a shell of thickness $2\delta$
in the limit $\delta \to 0.$ Derivatives of quantities discontinuous
across the brane will generate singular distributions on the brane
which relate to the localization of the stress-energy.
This relation is encompassed by the matching conditions across the
brane obtained by the integration of the corresponding equation of
motion in the direction normal to the brane. The matching conditions
provide the boundary conditions on the brane for the bulk fields, thus
constraining the parallel projected equations to produce the induced
equations on the brane.
Spontaneous symmetry breaking is then treated by assuming that the
bulk vector field acquires a non-vanishing expectation value which
reflects on the brane the breaking of the Lorentz symmetry in the bulk.

\section{Bulk Vector Field Coupled to Gravity}

Aiming to examine the gravitational effects of the breaking of Lorentz
symmetry in a braneworld scenario,
we consider a bulk vector field ${\bf B}$ with a non-minimal coupling
to the graviton in a five-dimensional anti-de Sitter space.
The Lagrangian density consists of
the Hilbert term, the cosmological constant term,
the kinetic and potential terms for
${\bf B}$ and the ${\bf B}$--graviton interaction term, as follows
\begin{eqnarray}
{\cal L}
={1\over {\kappa_{(5)}^2}}R -2\Lambda
+\xi B^{\mu}B^{\nu}R_{\mu\nu}
-{1\over 4}B_{\mu\nu}B^{\mu\nu} -V(B^\mu B_\mu \pm b^2 ),\quad
\end{eqnarray}
where $B_{\mu\nu} =\nabla_{\mu}B_{\nu} -\nabla_{\nu}B_{\mu}$ is the
tensor field associated with $B_{\mu}$ and $V$ is the potential which
induces the spontaneous global symmetry breaking when the ${\bf B}$ field is
driven to the minimum at $B^\mu B_\mu \pm b^2 = 0 $, $b^2$ being
a real positive constant.
Here, $\kappa_{(5)}^2=8\pi G_{N}=M_{Pl}^3,$
$M_{Pl}$ is the five-dimensional Planck mass
and $\xi$ is a dimensionless coupling constant that we have
inserted to track the effect of the interaction.
In the cosmological constant term
$\Lambda =\Lambda_{(5)} +\Lambda_{(4)}$ we have included both the bulk
vacuum value $\Lambda_{(5)}$ and that of the brane $\Lambda_{(4)},$
described by a brane tension $\sigma$ localized on the locus of the
brane, $\Lambda_{(4)}=\sigma\delta(N).$

The Einstein equation is given by:
\begin{eqnarray}
{1\over \kappa_{(5)}^2}G_{\mu\nu} +\Lambda g_{\mu\nu}
-\xi L_{\mu\nu} -\xi \Sigma_{\mu\nu} ={1\over 2}T_{\mu\nu},
\end{eqnarray}
where
\begin{eqnarray}
L_{\mu\nu} &=&{1\over 2}g_{\mu\nu}B^{\rho}B^{\sigma}R_{\rho\sigma}
-\left( B_{\mu}B^{\rho}R_{\rho\nu} +R_{\mu\rho}B^{\rho}B_{\nu}\right),\qquad\\
\Sigma_{\mu\nu} &=& {1\over 2}\bigl[
\nabla_{\mu}\nabla_{\rho}(B_{\nu}B^{\rho})
+\nabla_{\nu}\nabla_{\rho}(B_{\mu}B^{\rho})\
-\nabla^2(B_{\mu}B_{\nu})
-g_{\mu\nu}\nabla_{\rho}\nabla_{\sigma}(B^{\rho}B^{\sigma})\bigr]
\end{eqnarray}
are the contributions from the interaction term and
\begin{eqnarray}
T_{\mu\nu} =
B_{\mu\rho}B_{\nu}{}^{\rho} +4V^\prime B_{\mu}B_{\nu}
+g_{\mu\nu}\left[ -{1\over 4}B_{\rho\sigma}B^{\rho\sigma} -V\right] \quad
\end{eqnarray}
is the contribution from the vector field for the stress-energy tensor.
For the equation of motion for the vector field ${\bf B}$, we find that
\begin{eqnarray}
\nabla^{\nu}
 \left( \nabla_{\nu}B_{\mu} -\nabla_{\mu}B_{\nu}\right)
-2V^{\prime}B_{\mu}
+2\xi B^{\nu}R_{\mu\nu}
=0. \quad
\end{eqnarray}
where $V^{\prime} =dV/dB^2$. Projecting the equations parallel (A) and orthogonal (N) to
the surface of the brane, we proceeded to integrate them in the normal direction to the
extract the matching conditions. These conditions constrain the parallel projected
equations to yield the induced equations on the brane.  The general features of this
procedure have been previously discussed\cite{BuCarvalho05}.

When the bulk vector field ${\bf B}$ acquires a non-vanishing,
covariantly conserved\cite{Kostelecky3} vacuum expectation value by
spontaneous symmetry breaking, the bulk vacuum acquires an intrinsic
direction determined by $\left<B_{A}\right>,$ thus inducing the
breaking of the Lorentz symmetry in the bulk.
In order to obtain a vanishing cosmological constant
and ensure that Lorentz invariance holds on the brane, we take the
Einstein equation induced on the brane and impose
respectively that
\begin{equation}
\Lambda_{(5)} = {1\over 2}(1 -2(\xi -1))K\sigma
\label{eqn:vanishingcc}
\end{equation}
and that
\begin{eqnarray}
&&{1\over \kappa_{(5)} ^2}\biggl[
2K_{AC}K_{BC}
-\left( {1\over 2} +\xi -1\right)K_{AB}K \cr
&&+{1\over 2}g_{AB}\left(
R^{(ind)} -2K_{CD}K_{CD}
-\left( 1 -2(\xi -1)\right)K^2\right)
\biggr]
\cr
&=&
{\xi\over 2}\biggl[
\left( {5\over 2} -2 +{2\over \xi}\right)\left(
\left<B_{A}\right>\left<B_{C}\right>R^{(ind)}_{CB}
+\left<B_{B}\right>\left<B_{C}\right>R^{(ind)}_{AC}\right)
-(4\xi +2)K_{AC}K_{BD}\left< B_{C}\right>\left< B_{D}\right>
\biggr]\cr
&+&{\xi\over 2}g_{AB}\biggl[
\left< B_{C}\right>\left< B_{D}\right>R^{(ind)}_{CD}
+2(\xi -1)K_{CE}K_{ED}\left< B_{C}\right>\left< B_{D}\right>\biggr]
\label{eqn:Lorentzbrane},
\end{eqnarray}
which for $\xi =1$ reduce to the results presented in [2].

\section{Discussion and Conclusions}

In this contribution we examine the spontaneous symmetry breaking of Lorentz invariance in
the bulk and its effects on the brane. For this purpose, we considered a bulk vector field
subject to a potential which endows the field with a non-vanishing vacuum expectation
value, thus allowing for the spontaneous breaking of the Lorentz symmetry in the bulk.
This bulk vector field is directly coupled to the Ricci tensor so that, after the breaking
of Lorentz invariance, the breaking of this symmetry is transmitted to the gravitational
sector. We assign a non-vanishing vacuum expectation value to the component of ${\bf B}$
parallel to the brane (the generality of this procedure has been discussed in [8]). We
observe that there is a connection between the vanishing of the cosmological constant and
the reproduction of the Lorentz invariance on the brane. The conditions above were
enforced so that the higher dimensional signatures encapsulated in the induced geometry of
the brane cancel the Lorentz symmetry breaking inevitably induced on the brane, thus
reproducing the observed geometry. Naturally, the first condition,
Eq.~(\ref{eqn:vanishingcc}), can be modified to account for any non-vanishing value for
the cosmological constant induced on the brane. A much more elaborate fine-tuning,
however, is required for the Lorentz symmetry to be observed on the brane, as expressed by
the condition Eq.~(\ref{eqn:Lorentzbrane}). We believe that this is a new feature in
braneworld models, as in most such models Lorentz invariance is a symmetry shared by both
the bulk and the brane. Notice that a connection between the cosmological constant and
Lorentz symmetry had been conjectured long ago\cite{Bertolami3}. We shall examine further
implications of this mechanism in a forthcoming publication where we will also discuss the
inclusion of a bulk scalar field \cite{BeCarvalho07}.


\end{document}